\documentclass{jpp}
\usepackage{graphicx}
\usepackage{epstopdf, epsfig}
\usepackage{hyperref}
\usepackage{amsmath,amsfonts,amssymb}
\usepackage{subfigure}

\newcommand{\pa}[2]{\frac{\partial #1}{\partial #2}}
\newcommand{\paf}[2]{\partial #1/\partial #2}
\newcommand{\ve}[1]{\mathbf{#1}}

\graphicspath{{figs}}

\shorttitle{Restoring momentum conservation to magnetized quasilinear diffusion}
\shortauthor{I. E. Ochs}

\title{Restoring momentum conservation to magnetized quasilinear diffusion}

\author{I. E. Ochs\aff{1}\corresp{\email{iochs@princeton.edu}}}

\affiliation{\aff{1}Department of Astrophysical Sciences, Princeton University, Princeton, NJ, USA}

\begin{document}

\maketitle

\begin{abstract}
Wave interactions with magnetized particles underly many plasma heating and current drive technologies.
Typically, these interactions are modeled by bounce-averaging the quasilinear Kennel-Engelmann diffusion tensor over the particle orbit.
However, as an object derived in a two-dimensional space, the Kennel-Engelmann tensor does not fully respect the conservation of four-momentum required by the action conservation theorem, since it neglects the absorption of perpendicular momentum.
This defect leads to incorrect predictions for the wave-induced cross-field particle transport.
Here, we show how this defect can easily be fixed, by extending the tensor from two to four dimensions and matching the form required by four-momentum conservation.
The resulting extended tensor, when bounce-averaged, recovers the form of the diffusion paths required by action-angle Hamiltonian theory.
Importantly, the extended tensor should be easily implementable in Fokker-Planck codes through a mild modification of the existing Kennel-Engelmann tensor.
\end{abstract}


\section{Introduction}

Wave-induced diffusion is an important tool for plasma confinement, allowing for both current drive \citep{Fisch1978ConfiningTokamak,Fisch1980CreatingAsymmetric,Fisch1987TheoryCurrent} and heating \citep{Adam1987ReviewTokamak,Thumm2019HighpowerGyrotrons} of fusion plasmas.
As a result of these wave-particle interactions, the particle orbit changes, leading to wave-induced transport across magnetic field lines and flux surfaces.
This cross-field transport is fundamentally linked to momentum exchange between the particle and the wave \citep{Lee2012PerpendicularMomentum,Guan2013PerpendicularMomentum,Guan2013PlasmaRotation,Guan2013ToroidalPlasma,Ochs2021WavedrivenTorques,Ochs2022MomentumConservation,Ochs2024WhenWaves}.
This combination of energetic and cross-field transport can drive instabilities, but can also enable beneficial effects such as alpha channeling, which aims to extract high-energy fusion-born ash while harvesting its energy for use in useful waves \citep{Fisch1992InteractionEnergetic,Fisch1992CurrentDrive,Valeo1994ExcitationLargekTheta,Herrmann1997CoolingEnergetic}.
Correctly modeling these processes requires a quasilinear diffusion theory that respects momentum conservation.

Historically, there have been two classes of approaches for modeling quasilinear wave-particle interactions.
In perhaps the most elegant approach, one begins with a global electromagnetic field Fourier-decomposed in action-angle coordinates, and then performs the quasilinear average in these coordinates \citep{Kaufman1972QuasilinearDiffusion,Eriksson1994MonteCarlo,Herrmann1998CoolingAlpha,Brizard2022HamiltonianFormulations}.
Because of the simple particle dynamics in these coordinates, the resulting expressions are fairly compact.
However, the Fourier-decomposition of even a simple wavefield can be quite complex, and thus this approach is not necessarily ideal for all situations, such as modern simulations where the background fields (and thus the mapping to the action-angle coordinates) can change, or those which couple to simulations of the wave field.

In the second approach, one calculates the quasilinear diffusion tensor locally according to the local wavevectors, making use of the straight-field-line Kennel-Engelmann diffusion tensor form \citep{Kennel1966VelocitySpace,Lerche1968QuasilinearTheory}.
Then, one bounce-averages this diffusion tensor over the particle trajectory.
This has been the standard approach to Fokker-Planck calculations of wave-particle interactions in mirror \citep{Bernstein1981RelativisticTheory,Matsuda1986RelativisticMultiregion,Frank2025ConfinementPerformance} and tokamak \citep{Harvey1992CQL3DFokkerplanck,Petrov2016FullyneoclassicalFiniteorbitwidth} plasmas, due to its relative simplicity and computational efficiency.

In order for the latter approach to accurately model wave-induced transport, one must start with a diffusion tensor that respects momentum conservation.
Unfortunately, in its conventional form, the Kennel-Engelmann tensor does not, as it satisfies only two of the four requisite momentum-conserving relations.

To proceed, we first review the momentum conservation relations that must be satisfied during a resonant quasilinear wave-particle interaction, reviewing how these result in cross-field diffusion in a magnetized plasma.
We also show that these imply a specific form for the diffusion tensor in energy-momentum space.
We then review the Kennel-Engelmann diffusion tensor, showing how it respects a subset of the energy-momentum conservation relations.
Using our understanding the required form of the diffusion tensor, we then trivially construct an extended Kennel-Engelman tensor that fully respects energy- and momentum-conservation.
Finally, we show how this new form of the tensor affects the form of constants-of-motion space quasilinear diffusion, in particular recovering the diffusion paths from the action-angle quasilinear theory.
Since it relies on the same underlying bounce-averaged quasilinear theory, this extended Kennel-Engelmann tensor should be able to be rapidly adopted into existing Fokker-Planck codes.

\section{Generic form of local diffusion tensor for a resonant wave-particle interaction}

Consider a particle, possibly in the presence of a uniform magnetic field.
Such a particle has a constant kinetic energy $K$ and canonical momentum components $\ve{p}$, which are often combined in the energy-momentum four-vector $p^\mu \equiv (K,\ve{p})$.

To this system, add a monchromatic wave (oscillating electromagnetic field) with frequency $\omega$ and wavevector $\ve{k}$.
These objects can also be combined into a wave four-vector $k^\mu \equiv (\omega, k^x, k^y, k^z)$.
Then, allow this wave to resonantly interact with the plasma, either through a Landau or gyro resonance.

It is well known \citep{Trubnikov1979,Stix1992WavesPlasmas,Dodin2012AxiomaticGeometrical} that, during the resonant wave-particle interaction, the increments of energy and momentum satisfy the relation:
\begin{align}
	\frac{\Delta p^\mu}{k^\mu} = \frac{\Delta p^\nu}{k^\nu}, \quad \forall \mu,\nu \in \{0,1,2,3\}. \label{eq:DiffusionPathDeltaFormulation}
\end{align}
This relationship follows from the conservation of the wave action $\mathcal{I}$, the definition of the Minkowski four-momentum $\mathcal{P}^\mu \equiv k^\mu \mathcal{I}$, and the fact that the Minkowski subsystem and resonant particle subsystem form a conserving system \citep{Ochs2021WavedrivenTorques,Ochs2022MomentumConservation,Ochs2024WhenWaves}.

\begin{figure}
	\centering
	\includegraphics[width=0.8\linewidth]{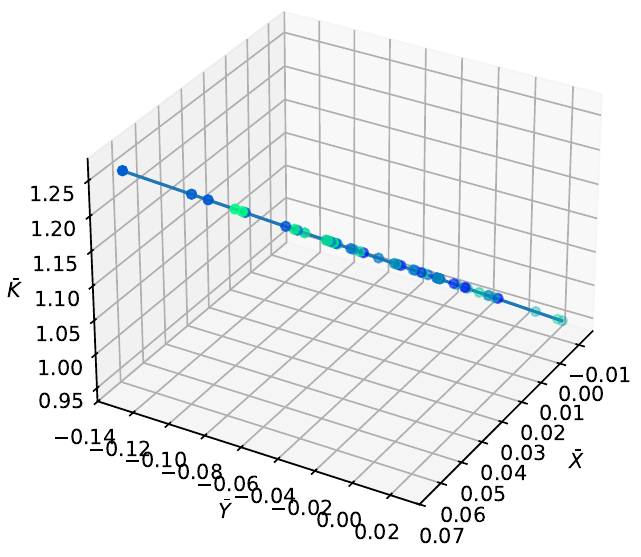}
	\caption{Simulation of ion interaction with a second-harmonic left-hand-polarized wave over many orbits. The particle diffuses in both normalized energy $\bar{K}$ and gyrocenter position $(\bar{X},\bar{Y})$ [points, with time corresponding to color], but stays on the line determined by Eq.~(\ref{eq:DiffusionPathDeltaFormulation}) [solid line].
		The normalized variables and details of the simulation can be found in Appendix~\ref{sec:SimDescription}.}
	\label{fig:DiffusionPath}
\end{figure}

In a magnetized system, the relation in Eq.~(\ref{eq:DiffusionPathDeltaFormulation}) leads to coupled gyrocenter-energy diffusion, as the change in the gyrocenter position is given by:
\begin{align}
	\Delta \ve{X} = \frac{\Delta \ve{p} \times \ve{B}}{B^2}. \label{eq:GyrocenterMomentum}
\end{align}
In other words, diffusion takes place along single line in gyrocenter-energy space---which is the basis for the alpha channeling effect \citep{Fisch1992InteractionEnergetic,Fisch1992CurrentDrive}.
Although the early references focused on diffusion due to high-frequency electrostatic waves, in fact the relation holds very generally for any electromagnetic wave.
For instance, Fig.~\ref{fig:DiffusionPath} shows diffusion in energy and gyrocenter space for a particle undergoing diffusion due to interaction with a left-polarized electromagnetic wave at the second cyclotron harmonic of a $\hat{z}$ directed magnetic field.
It can be seen that over many successive interactions with the wave, the particle stays on the line defined by Eqs.~(\ref{eq:DiffusionPathDeltaFormulation}-\ref{eq:GyrocenterMomentum}).
[The simulation is described in more detail in Appendix~\ref{sec:SimDescription}.]

The relation in Eq.~(\ref{eq:DiffusionPathDeltaFormulation}) means that the diffusion operator in four momentum space must take a very specific form, namely:
\begin{align}
	D^{\mu \nu} &= D^{K K} \frac{k^\mu k^\nu}{\omega^2}. \label{eq:DGenericForm}
\end{align}
Here, $D^{KK}$ is the on-diagonal kinetic-energy component of the tensor.
Eq.~(\ref{eq:DGenericForm}) is very powerful, because it allows us to construct a four-momentum-conserving wave-particle interaction tensor from a single known tensor component.

\section{Restoring conservation to the Kennel-Engelmann diffusion tensor}

Quasilinear diffusion in due to a monochromatic wave in a uniform magnetic field is traditionally modeled by the equation \citep{Kennel1966VelocitySpace,Lerche1968QuasilinearTheory,Stix1992WavesPlasmas,Harvey1992CQL3DFokkerplanck,Petrov2016FullyneoclassicalFiniteorbitwidth}:
\begin{align}
	\pa{f}{t} &= \frac{1}{\sqrt{g}} \pa{}{\ve{p}} \cdot \left(\sqrt{g} \, \ve{D} \cdot \pa{f}{\ve{p}}\right) \label{eq:QLDiffEq}
\end{align}
where the volume element $\sqrt{g} = 2\pi p_\perp$, and $\ve{D}$ is given by the Kennel-Engelmann diffusion tensor, which can be written
\begin{align}
	\ve{D} &= D_{0} \ve{w} \ve{w} \label{eq:DKE}\\
	D_{0} &= \frac{\pi q^2}{2} \sum_n | \psi_{n} |^2 \delta(\omega_{r} - k_\parallel v_\parallel - n\Omega) \label{eq:D0}\\
	\ve{w} &= \left(1 - \frac{k_\parallel v_\parallel}{\omega_{r}} \right) \hat{p}_\perp + \left(\frac{k_\parallel v_\perp}{\omega_{r}}\right) \hat{p}_\parallel. \label{eq:wDef}
\end{align}
Here,
\begin{align}
	\psi_{n} &\equiv E^+ J_{n-1}(z) + E^-  J_{n+1}(z) + \frac{p_\parallel}{p_\perp} E^z J_n(z),
\end{align}
where
\begin{align}
	z = \frac{k_\perp v_\perp}{\Omega}
\end{align}
and
\begin{align}
	E^\pm \equiv \frac{1}{2} \left(E_x \pm i E_y \right) e^{\mp i \theta},
\end{align}
where it has been assumed that $\ve{B} \parallel \hat{z}$, and where $\theta$ is the angle defined by:
\begin{align}
	\ve{k} &= \hat{x} k_\perp \cos \theta + \hat{y} k_\perp \sin \theta + \hat{z} k_\parallel.
\end{align}
[Note: the additional phase $i\theta$ is not included in the definition of $E^\pm$ in references, such as those for the CQL3D code \citep{Harvey1992CQL3DFokkerplanck,Petrov2016FullyneoclassicalFiniteorbitwidth}, which employ the ``Stix frame'', where $\ve{k}$ lies in the $x-z$ plane and thus $\theta = 0$.]

Eq.~(\ref{eq:QLDiffEq}) describes a diffusion process in two coordinates: the parallel momentum $p_\parallel \equiv p_z$, and the perpendicular momentum $p_\perp \equiv m v_\perp$.
It is important to note that the perpendicular momentum so defined is more precisely an energy variable, having nothing to do with the gyrocenter position.
Living in this reduced space, the Kennel-Engelmann diffusion tensor fundamentally cannot capture the full set of conservation relations demanded by Eq.~(\ref{eq:DiffusionPathDeltaFormulation}).

Nevertheless, the Kennel-Engelmann tensor does capture the energy-momentum relation in the dimensions that it does model.
To see this, we can transform the tensor from coordinates $(p_\perp,p_\parallel)$ to coordinates $(K,p_\parallel)$.
To do this, it is sufficient to transform the vector $\ve{w}$ [Eq.~(\ref{eq:wDef})].
Using 
\begin{align}
	K = \frac{1}{2m} \left(p_\perp^2 + p_\parallel^2\right) \Rightarrow \pa{K}{p_\perp} = \frac{p_\perp}{m}; \; \pa{K}{p_\parallel} = \frac{p_\parallel}{m}.
\end{align}
we find
\begin{align}
	w^K = \pa{K}{p_\perp} w^{p_\perp} + \pa{K}{p_\parallel} w^{p_\parallel} = v_\perp.
\end{align}
Comparing to Eq.~(\ref{eq:wDef}), we see that this satisfies:
\begin{align}
	w^K &= \frac{k_\parallel}{\omega} w^{p_\parallel}, \label{eq:wRelation}
\end{align}
implying [via Eq.~(\ref{eq:DKE})] that the diffusion satisfies the requirement in Eq.~(\ref{eq:DGenericForm}) for $K$ and $p_\parallel$.
Indeed, the connection of the Kennel-Engelmann diffusion path to the absorption of photon four-momentum has long been appreciated \citep{Kennel1966VelocitySpace,Stix1992WavesPlasmas}.

To make a momentum-conserving tensor is now trivial.
Noting that 
\begin{align}
	D^{K K} &= D_{0} v_\perp^2,
\end{align}
with $D_0$ given by Eq.~(\ref{eq:D0}), Eq.~(\ref{eq:DGenericForm}) then tells us that in four-momentum space $(K,p^x,p^y,p^z)$:
\begin{align}
	D^{\mu \nu} &= D_0 \frac{k^\mu k^\nu v_\perp^2}{\omega^2}.
\end{align}
The projection of this tensor to $(K,p_\parallel)$ is precisely the Kennel-Engelmann tensor, while the new components capture the absorption of perpendicular momentum.

\section{Constants-of-motion space diffusion}

A typical application of quasilinear diffusion theory is to bounce-averaged Fokker-Planck theory.
Such theories describe diffusion in the space of particle orbits, which can be specified by the constants of motion (COM) of the orbit.
It is thus instructive to examine the effects of the new terms in the diffusion tensor on the particle diffusion in COM space.

A typical choice of the COM in an axisymmetric system is $(\epsilon,\mu,p_\phi)$, where:
\begin{align}
	\epsilon &\equiv \frac{1}{2m} (p_\perp^2 + p_\parallel^2) + \psi(\ve{r})\\
	\mu &\equiv \frac{p_\perp^2}{2m |B|}\\
	p_\phi &\equiv (\ve{r} \times \ve{p}) \cdot \hat{z} + q r A_\phi(r).
\end{align}
Here, $\ve{r} \equiv (r,\phi,z)$ is the cylindrical coordinate vector, $\psi(\ve{r})$ is a position-dependent potential energy, and $A_\phi(\ve{r})$ is the $\phi$-component of the vector potential.
The $p_\phi$ component is closely related to the particle flux surface, and is often taken to define the flux surface.
In these coordinates, the diffusion tensor again has the form from Eq.~(\ref{eq:DKE}), and the task is to calculate the diffusion path vector components $w^\epsilon$, $w^\mu$, and $w^{p_\phi}$.

It is clear that $w^\epsilon$ and $w^\mu$ are unchanged by the new momentum-conserving terms in the diffusion tensor.
However, $w^{p_\phi}$ is certainly changed, as:
\begin{align}
	w^{p_\phi} &= \pa{p_\phi}{p_i} w^{p_i}, \label{eq:wpphi1}
\end{align}
which involves a sum over the $\phi$-projection of all three momentum components, rather than just the parallel momentum.
In a linear-confinement axisymmetric system such as a magnetic mirror, where $\paf{p_\phi}{p_\parallel} = 0$, the difference to the traditional Kennel-Engelmann theory is particularly striking, as the new components make the difference between nonexistent cross-flux-surface transport, versus cross-flux-surface transport due to perpendicular momentum absorption.

Furthermore, we can see easily that the new terms make good sense.
Plugging in $w^{p_i} = k^i v_\perp / \omega$ [cp. Eqs.~(\ref{eq:DGenericForm}), (\ref{eq:DKE}) and (\ref{eq:wRelation})] to Eq.~(\ref{eq:wpphi1}), we find:
\begin{align}
	w^{p_\phi} = \left(\ve{r} \times \frac{\ve{k} v_\perp}{\omega} \right) \cdot \hat{z} = \frac{\ve{r} \times \ve{k}\cdot \hat{z}}{\omega} w^{K} , \label{eq:wpphi2}
\end{align}
the natural relation between the absorbed photon energy and angular momentum.
Thus, we see that the change in the diffusion tensor modifies the momentum absorption so as to enforce angular momentum conservation.

For completeness, we can now fully derive the form of the diffusion tensor in COM space.
It is clear that $w^\epsilon = w^K$.
Eq.~(\ref{eq:wpphi2}) can be expressed in more familiar form by making use of the azimuthal mode number of the wave $n_\phi$, from whence 
\begin{align}
	w^{p_\phi} = \frac{n_\phi}{\omega} w^\epsilon.
\end{align}
Then the final task is to find $w^\mu$.
Noting that 
\begin{align}
	\mu = \frac{1}{|B|} \left(K - \frac{1}{2m} p_\parallel^2 \right)
\end{align}
and making use of the $\delta$-function in Eq.~(\ref{eq:D0}), we find:
\begin{align}
	w^{\mu} &= \frac{1}{|B|} \left( 1 - \frac{k_\parallel v_\parallel}{\omega} \right) w^\epsilon = \frac{n \Omega}{|B| \omega} w^{\epsilon}.
\end{align}
Thus, in $(\epsilon,\mu,p_\phi)$ space, the diffusion tensor is given by Eq.~(\ref{eq:DKE}), but with:
\begin{align}
	\ve{w} &= v_\perp \begin{pmatrix} 1 \\ \frac{n \Omega}{|B| \omega} \\ \frac{n_\phi}{\omega} \end{pmatrix}.
\end{align}
This has the same form of the diffusion path found in the action-angle space quasilinear theory [see e.g. \cite{Herrmann1998CoolingAlpha} Eq.~(4.67)], but now calculated with a diffusion coefficient from the conventional bounce-averaged theory.

\section{Discussion}

In this paper, we have shown how to extend the Kennel-Engelmann diffusion tensor to respect quasilinear four-momentum conservation.
By converting the resulting tensor to axisymmetric constants-of-motion space, we have shown that the extended Kennel-Engelmann tensor modifies the cross-field wave-induced transport, recovering the diffusion path from the action-angle coordinate Hamiltonian theory \citep{Herrmann1997CoolingEnergetic,Herrmann1998CoolingAlpha}.
Thus, past wave transport calculations that did not rely on bounce-averaging the 2D diffusion tensor, are unaffected by the new form of the diffusion tensor; this category includes alpha channeling theories derived by reconstructing diffusion paths in gyrocenter-energy space \citep{Fisch1992InteractionEnergetic,Fisch1992CurrentDrive,Valeo1994ExcitationLargekTheta,Ochs2015CouplingAlpha,Ochs2015AlphaChanneling,Romanelli2020InteractionIon}, or derived using action-angle quasilinear theory  in toroidal coordinates \citep{Kaufman1972QuasilinearDiffusion,Eriksson1994MonteCarlo,Herrmann1997CoolingEnergetic,Herrmann1998CoolingAlpha}.
Instead, the current work allows the correction of calculations of wave-induced cross-field transport in calculations and codes that rely on bounce-averaging the 2D Kennel-Engelmann diffusion operator, such as CQL3D \citep{Harvey1992CQL3DFokkerplanck,Petrov2016FullyneoclassicalFiniteorbitwidth}.
Fortunately, the formulation here allows such codes to be quickly updated by adding a few terms to the existing operators.
Such improvements are important in any geometry, but are likely to be particularly critical for transport in magnetic mirrors, where radial transport is driven exclusively by the absorption of perpendicular momentum.

It should be emphasized that the extended Kennel-Engelmann diffusion tensor derived here only describes the action of the wave on resonant particles. 
To get the total cross-field transport resulting from the wave, one must include nonresonant effects such as the ponderomotive recoil \citep{Ochs2021NonresonantDiffusion,Ochs2023PonderomotiveRecoil}, which can also drive cross-field transport. In general, this distinction is only an issue for time-dependent wave envelopes; if the wave envelope is stationary, then the resonant diffusion represents the total cross-field transport due to the wave \citep{Ochs2021WavedrivenTorques,Ochs2022MomentumConservation,Ochs2024WhenWaves}.

Finally, we note here that demanding the tensor form in Eq.~(\ref{eq:DGenericForm}) has the potential to ease the calculation of other resonant diffusion tensors, since the entire tensor can be constructed from a single component.
Consider, for instance, the non-Bessel-function form of the magnetized susceptibility \citep{Qin2007NewDerivation}, which should allow for asymptotic expansion of the quasilinear diffusion tensor in the large-gyroradius (large-$z$) limit.
Rather than performing the asymptotic expansion on a two-dimensional diffusion tensor, the approach taken here allows one to expand a single component, and then end up with a four-dimensional momentum-conserving tensor that includes the induced cross-field transport. 
Thus, leveraging the energy and momentum conservation allows one to produce a more powerful object with much less algebra.

\section*{Acknowledgements}

The author would like to thank Nat Fisch for useful discussions.
This work was partially supported by Department of Energy Grant No. DE-SC0016072, and partially by Princeton University. 
%
%

\appendix

\section{Description of simulations} \label{sec:SimDescription}

The simulations in the main text use a second-order modification to the Boris algorithm \citep{Zenitani2018BorisSolver} to  solve the Lorentz force equations in normalized form:
\begin{align}
	\frac{d\ve{\bar{v}}}{d\bar{t}} &= \ve{\bar{E}} + \ve{\bar{v}} \times \ve{\bar{\Omega}};\\
	\frac{d\ve{\bar{x}}}{d\bar{t}} &= \ve{\bar{v}}.
\end{align}
Here, $t$ is normalized to $\Omega_0^{-1}$, and the electric field is normalized to $E_0 = B_0 v_0 /c$, where $v_0$ is an arbitrarily normalized velocity.

We take a system with a primarily $\hat{z}$-directed uniform background magnetic field, with a small variation along $\hat{z}$:
\begin{align}
	\ve{\bar{\Omega}} = \left[1 + \Delta \sin \left(\frac{2 \pi z}{L} \right)\right]\hat{z} - \pi \frac{\bar{r}}{L} \Delta \cos \left(\frac{2 \pi z}{L} \right) \hat{r}.
\end{align}

For a left-hand polarized wave, we take the vector potential:
\begin{align}
	\ve{\bar{A}} &= \frac{\bar{E}_{w0}}{\bar{\omega}} e^{-\bar{z}_m^2/a^2} \left(\sin \bar{\theta} \hat{x} - \cos \bar{\theta} \hat{y} \right),\\
	\bar{\theta} &\equiv \ve{\bar{k}} \cdot \ve{\bar{x}} - \bar{\omega} \bar{t} + \bar{\phi},
\end{align}
from whence
\begin{align}
	\ve{\bar{E}} = -\pa{\ve{\bar{A}}}{\bar{t}}; \quad \ve{\bar{B}} = \bar{\nabla} \times \ve{\bar{A}}.
\end{align}
Here, $\bar{z}_m = \mod (z + L/2,L) - L/2$.
The length $a$ is taken to be much shorter than the system periodicity $L$, so that the wave essentially vanishes at the system ends.
The phase $\bar{\phi}$ is generally constant, but is randomized once per orbit as the particle passes the point furthest from the wave envelope.

\begin{figure}
	\centering
	\includegraphics[width=0.8\linewidth]{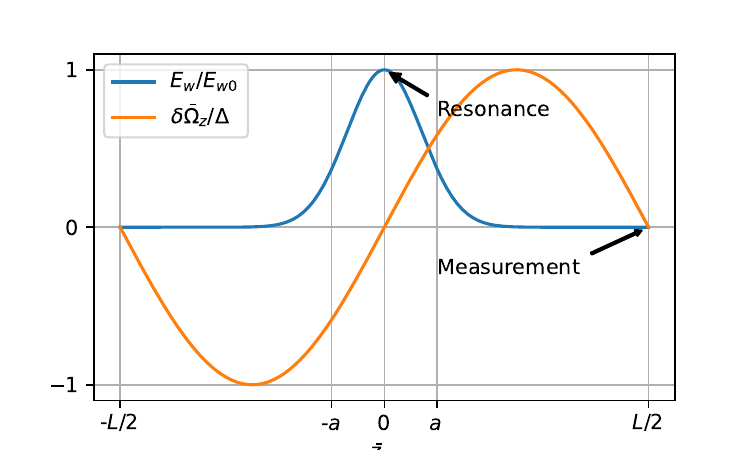}
	\caption{Amplitude of normalized electric field and axial magnetic field in the simulation.
		The resonant interaction occurs around $z_m = 0$; the measurements of gyrocenter and energy are taken arund $z_m = L/2$.}
	\label{fig:SimulationSchematic}
\end{figure}

The logic of the above system is as follows.
The magnetic field adopts its mean value twice per length $L$: once at $z_m = 0$, and once at $z_m = L/2$.
At $z_m = 0$, we place the wave field, which is resonant exactly at $z_m = 0$ (as we take $\bar{k}_z = 0$ and $\bar{\omega} \in \mathbb{Z}$). 
Then, we cleanly measure the gyrocenter position and energy at $z_m = L/2$, far from the interfering wave field.
The phase randomization preserves the random phase approximation inherent in the quasilinear theory, preventing coherent structure formation due to resonances between the bounce and wave motion.

The code outputs the gyrocenter position $\ve{\bar{X}}$, and the parallel and perpendicular velocities $\bar{v}_\parallel$ and $\bar{v}_\perp$ in the gyrocenter drift frame.
The dimensionless energy is given by $\bar{K} = \frac{1}{2} \left(\bar{v}_\perp^2 + \bar{v}_\parallel^2 \right)$.
The gyrocenter is related to the canonical momentum by:
\begin{align}
	\Delta \ve{\bar{X}} &= \Delta \ve{\bar{p}} \times \ve{\bar{\Omega}}. \label{eq:DeltaXgcApp}
\end{align}
Combining Eq.~(\ref{eq:DeltaXgcApp}) with Eq.~(\ref{eq:DiffusionPathDeltaFormulation}), the diffusion should occur on the line:
\begin{align}
	\bar{\omega} \Delta \bar{K} = - \bar{k}_x \Delta \bar{Y} = \bar{k}_y \Delta \bar{X}.
\end{align}
This is the line shown in Fig.~\ref{fig:DiffusionPath}.

For the simulations shown in Fig.~\ref{fig:DiffusionPath}, the parameters are: $\Delta = 0.07$, $L = 2000$, $a = 50$, $\bar{E}_{w0} = 0.0015$, $\bar{\omega} = 2$, $\bar{k}_x = 1$, $\bar{k}_y = 0.5$.
The particle was initialized with $\bar{v}_\perp = \bar{v}_\parallel = 1$, and followed for 50 transits of the system.


\clearpage
\newpage

\end{document}